\documentclass[12pt]{iopart}

\usepackage{setstack}
\usepackage{bm}
\usepackage{graphicx}
\usepackage{amsbsy}
\usepackage{amstext}
\usepackage{amssymb}
\usepackage{units}
\usepackage{hyperref}
\usepackage{xcolor}

\begin{document}

\title[Magnetic oscillations in graphene with spin splitting: a new approach]{Influence of temperature on the magnetic oscillations in graphene with spin splitting: a new approach}

\author{F Escudero, J S Ardenghi,
	P Jasen}

\address{IFISUR, Departamento de F\'isica (UNS-CONICET), Av. Alem 1253, B8000CPB Bah\'ia Blanca, Argentina\\
	Instituto de F\'isica del Sur (IFISUR, UNS-CONICET)\\
	Av. Alem 1253, B8000CPB Bah\'ia Blanca, Argentina}
\ead{federico.escudero@uns.edu.ar}

\begin{abstract}
We analyze the magnetic oscillations (MO) due to the de Haas-van Alphen effect, in pristine graphene under a perpendicular magnetic field, taking into account the Zeeman effect. We consider a constant Fermi energy, such that the valence band is always full and only the conduction band is available. At zero temperature the MO consist of two sawtooth peaks, one for each spin. Both peaks have the same frequency, but different amplitude and phase. We show that, in order to observe the spin splitting in the MO, Fermi energy of about 0.1 eV is required. At low temperatures we obtain that the MO can be expressed as the MO at zero temperature, plus small Fermi-Dirac like functions, each centered around the MO peaks. Using this expression, we show that the spin splitting is observable in the MO only when the thermal energy is smaller than the Zeeman energy. We also analyze the shift of the MO extrema as the temperature increases. We show that it depends on the magnetic field, which implies a broken periodicity at nonzero temperature. Finally, we obtain an analytical expression for the MO envelope. The results obtained could be used to infer temperature changes from the MO extrema shift and vice versa. 
\end{abstract}

%\keyword{graphene, magnetic oscillations, spin splitting}
\maketitle

\section{Introduction}

Since its experimental isolation in 2004, graphene has become one of the most interesting and promising materials in condensed matter science \cite{Novoselov2005,Geim2007,Zhang2005}. Its unique features, like high heat and electrical conductivity, has make graphene an exciting material for technological applications \cite{Avouris2012,Novoselov2012}. These properties are related to its 2D hexagonal structure, composed of two interpenetrating sublattices A and B \cite{Wallace1947}. Without impurities or defects, the conduction and
valence bands touch at the Fermi energy, with the valence band full
and the conduction band empty in the ground state \cite{Neto2009}.
Furthermore, in pristine graphene the density of states at the Fermi
energy is zero, making graphene a semiconductor with zero
band gap, or a semi-metal \cite{Lu2013}. In the long wavelength approximation
the dispersion relation is relativistic and the electrons behave as
massless fermions, moving with a Fermi velocity $\sim 10^6$ m/s \cite{Hwang2012}.

The magnetic properties of graphene and 2D systems have been investigated in recent works \cite{Goerbig2011,Ardenghi2015,Ardenghi2014,Ardenghi2013,Escudero2018}. Unlike conventional materials, the magnetization
in graphene has unique features. This can be related to the
Landau levels (LL) that appear when a magnetic field is applied \cite{Kuru2009}. In
a classical system, these LL are equidistant due to the parabolic
dispersion relation. But in graphene, at low energies the
dispersion relation is relativistic \cite{Neto2009}, causing the LL to be not
equidistant, which in turn affects the oscillating behavior in the
thermodynamics potentials \cite{Goerbig2011}. For instance, the magnetic oscillations
(MO), the so called de Haas van Alphen effect \cite{deHaas}, are sawtooth at
zero temperature \cite{Sharapov2004,Fu2011}, with the peaks being caused by the change in the last occupied
energy level \cite{Escudero2017}. 

The MO are altered by impurities and temperature. In each case the
MO are broadened as a result of the modified density of states (DOS)
and the Fermi-Dirac distribution. In classical metals this is described
by the Lifshitz-Kosevich formula \cite{Shoenberg1984}, which incorporates the effects of
impurities and temperatures, as well as the spin. The essential feature
of this formula is that the MO are written as an infinite series,
where the damping effects are taken into account by the introduction
of reduction factors. This formula has been extended to the case of
graphene \cite{Sharapov2004}, where the difference only lies in the form the reduction
factors take. In principle, this formula would be sufficient
to describe the MO in the most general case, but due to its complexity
it is not very useful. Thus, in general, limit cases are considered.
The most familiar one is zero temperature an energy independent impurities, where the only
reduction factor is the Dingle factor, in which case the infinite
series can be easily evaluated. Another interesting situation is the
pristine case (i.e. without impurities) at nonzero temperature. In this case the
general series cannot yet be evaluated, due to the complicated temperature
reducing factor, so in general some approximations are considered.
The most prominent one is the limit $B\rightarrow0$ (low magnetic
field), where one can simplify the temperature reduction factor and
easily evaluate the infinite series. Nevertheless, one could argue
that this limit may not be very useful, since the lower the magnetic
field the more difficult is to observe the MO. 

When the Zeeman effect is considered, a splitting of the Landau levels appears, which is relevant for several thermodynamical properties \cite{Escudero2017}. In turn, when a gate voltage is applied, the electron density can be changed, which could be useful for spin filtering \cite{Hanson2004}, spin-polarized currents in 2D systems \cite{Potok2003} and spin conductivity \cite{Sinitsyn2004}. Moreover, when other effects are considered, such as spin-orbit coupling due to an external electric field, the MO changes drastically \cite{Escudero2017a} and the interplay between the degeneracy of each level, the disorder and spin-orbit coupling is not trivial \cite{Escudero2017b}.

Motivated by this, we analyzed the MO in pristine graphene, in the presence of a perpendicular magnetic field, taking into account the Zeeman effect. We have organized this work as follow: in section 2 we study the MO in graphene at zero temperature, taking into the account the Zeeman effect. We then analyze the conditions to observe the spin splitting in the MO. In section 3 we study the MO at nonzero temperature, where we make an approximation to obtain an useful analytical formula for the magnetization. From this we analyze how the temperature affects the observation of the spin splitting in the MO. We end with a study of the maxima and minima shift in the MO as a function of the temperature, from which we obtain an expression for the oscillation envelope. Finally, our conclusions follow in section 4.

\section{MO at zero temperature}

In order to study the MO, we will consider that the Fermi energy $\mu$
is held constant. Moreover, we shall always take $\mu>0$, so that the valence band is always full. Given that $\mu$
is fixed, while the number of electrons $N$ may change, it is convenient
to work with the grand potential $\Omega$. We will consider the long
wavelength approximation, with energies close to the Fermi level.
This condition is satisfied as long as \cite{Neto2009} $\mu\ll\left|t\right|\sim3$
eV ($t$ is the nearest neighbor hopping amplitude). In this case the electrons
in graphene behave as relativistic massless fermions, whose dynamics
is given by the corresponding Dirac Hamiltonian. In the presence of
a perpendicular magnetic field $B$ one obtains the discrete Landau
level, as well as a spin splitting due to the Zeeman effect \cite{Zeeman1897}. Then
it can be shown \cite{Escudero2017} that in graphene the energy levels are given by $\varepsilon_{\lambda,n,s}=\lambda\alpha\sqrt{nB}-s\mu_{\mathrm{B}}B$,
where $\alpha=\upsilon_{F}\sqrt{2e\hbar}$ and $\upsilon_{F}\sim10^{6}$
m/s is the Fermi velocity, while $\lambda=\pm1$ for the conduction and
valence bands, $n=0,\:1,\,2,\ldots$ for the Landau level (LL) and
$s=\pm1$ for the spin (throughout this paper we will use +1 to indicate spin up and -1 to indicate spin down). Each energy level has a degeneracy
given by $D=2\mathcal{A}B/\phi$, where $\mathcal{A}$ is the sheet
area of graphene, $\phi=h/e$ is the magnetic unit flux and the factor
of 2 takes into account the valley degeneracy \cite{Goerbig2011}. 

For a Fermi energy $\mu>0$ at zero temperature, the valence band
is full while the conduction band is partially filled. We will write
the conduction energy levels as $\varepsilon_{m}=\alpha\sqrt{n_{m}B}-s_{m}\mu_{\mathrm{B}}B,$
where we have introduced the decreasing energy sorting index $m=0,\:1,\,2,\ldots$,
so $n_{m}$ gives the LL and $s_{m}$ the spin for the $m$ position.
In general, the mixing of the LL depends on the magnetic field \cite{Chang1997}, but for usual values
there is no spin mixing so  $n_{m}=\sqrt{m/2-\left[1-\left(-1\right)^{m}\right]/4}$
and $s_{m}=\left(-1\right)^{m}$. At a given $\mu>0$, all conduction energy
levels $m=0,\:1,\,2,\ldots,\,f$ are filled, where $f$ is such that
$\varepsilon_{f}<\mu\leq\varepsilon_{f+1}$. Then the grand potential
at zero temperature is

\begin{equation}
\Omega=\Omega_{V}+\sum_{m=0}^{f}D\left(\varepsilon_{m}-\mu\right),\label{GP}
\end{equation}
where $\Omega_{V}$ is the grand potential due to the filled valence
band. From equation (\ref{GP}) we shall expect the oscillatory contribution
coming only from the last term, associated with the change in the
last energy level as $B$ is modified. On the other hand, the first
term $\Omega_{V}$ makes a non-oscillatory contribution since the
valence band is always full if $\mu>0$. Therefore we omit for
the moment $\Omega_{V}$ and consider only the conduction grand potential
$\Omega_{C}=\sum_{m=0}^{f}D\left(\varepsilon_{m}-\mu\right)$. Separating
$\varepsilon_{m}=\varepsilon_{m}^{0}-s_{m}\mu_{\mathrm{B}}B$, with
$\varepsilon_{m}^{0}=\alpha\sqrt{n_{m}B}$, we get

\begin{equation}
\Omega_{C}=\Omega_{0}-D\mu_{\mathrm{B}}B\sum_{m=0}^{f}s_{m},\label{GP2}
\end{equation}
where $\Omega_{0}=\sum_{m=0}^{f}D\left(\varepsilon_{m}^{0}-\mu\right)$. 
The last term in equation (\ref{GP2}) is related to the Pauli paramagnetism $M_P$,
associated with the spin population. Indeed, we have $D\sum_{m=0}^{f}s_{m}=N_{+}-N_{-}$,
where $N_{+}$($N_{-}$) is the total number of spin up(down) states. Thus
$\mu_{\mathrm{B}}D\sum_{m=0}^{f}s_{m}=\mu_{\mathrm{B}}D\left[1+\left(-1\right)^{f}\right]/2=\mu_{\mathrm{B}}\left(N_{+}-N_{-}\right)=M_{p}$,
so equation (\ref{GP2}) becomes

\begin{equation}
\Omega_{C}=\Omega_{0}-BM_{P}.\label{GP3}
\end{equation}
The conduction magnetization is given by $M_{C}=-\mathcal{A}^{-1}\left(\partial\Omega_{C}/\partial B\right)_{\mu}$,
where $\mathcal{A}$ is the graphene area. From equation (\ref{GP3}) we
get 

\begin{equation}
M_{C}=M_{0}+\frac{1}{\mathcal{A}}\left(M_{P}+B\frac{\partial M_{P}}{\partial B}\right),\label{Mc-1}
\end{equation}
where $M_{0}=-\mathcal{A}^{-1}\left(\partial\Omega_{0}/\partial B\right)_{\mu}$.
Given that $\partial D/\partial B=D/B$, while $\partial\varepsilon_{m}^{0}/\partial B=\varepsilon_{m}^{0}/2B$,
we can write

\begin{equation}
M_{C}=-\frac{1}{2B}\left(\frac{3\Omega_{C}}{\mathcal{A}}+n\mu\right)+\frac{1}{2}m_{P},\label{Mc}
\end{equation}
where $n=N/\mathcal{A}=\sum_{m=0}^{f}D/\mathcal{A}=D\left(f-1\right)/\mathcal{A}$
is the density of conduction electrons and $m_{p}=M_{P}/\mathcal{A}$. Equation (\ref{Mc})
shows that the MO have a sawtooth oscillation (SO) produced whenever
$n$ or $m_{P}$ changes discontinuously, $\Omega_{C}$ being continuous
always. The sawtooth peaks amplitude $\Delta M$ are given by

\begin{equation}
\Delta M=-\frac{\mu}{2B}\Delta n+\frac{1}{2}\Delta m_{P}.\label{DeltaM}
\end{equation}
Each contribution $\Delta n$ and $\Delta m_{P}$ is determined
by the discontinuous change in the parameters $n_{f}$ and $s_{f}$,
which define the last energy level occupied. Considering the possible
changes of LL and spin, we can write the MO as a sum of two
sawtooth. In general the period of oscillation is given by $\Delta(1/B)=1/B_{2}-1/B_{1}$,
where $B_{i}$ is such that $\varepsilon_{f_{i}}(B_{i})=\mu$. Therefore
$\mu=\alpha\sqrt{B_{i}n_{i}}-s_{i}\mu_{\mathrm{B}}B_{i}$, where for simplicity
we have noted $n_{f_{i}}=n_{i}$ and $s_{f_{i}}=s_{i}$.
Using the approximation $\mu^{2}+2s_{i}\mu\mu_{\mathrm{B}}B_{i}+\mu_{\mathrm{B}}^{2}B_{i}^{2}\simeq\mu^{2}+2s_{i}\mu\mu_{\mathrm{B}}B_{i}$,
which holds for typical values of magnetic field, we obtain the $B_{i}$
at which the peaks occur:

\begin{equation}
\frac{1}{B_{i}}=\frac{n_{i}\alpha^{2}}{\mu^{2}}-\frac{2s_{i}\mu_{\mathrm{B}}}{\mu}.\label{1/Bi}
\end{equation}
We shall consider that the two peaks correspond to a fixed spin, with the oscillation being given when the LL changes by one.
Thus we take $\Delta n=n_{2}-n_{1}=1$, while $s=s_1=s_2$.
Therefore, from equation (\ref{1/Bi}) we obtain the period $\Delta(1/B)$
and frequency $\omega=\left[\Delta(1/B)\right]^{-1}$

\begin{equation}
\omega_{s}=\frac{\mu^{2}}{\alpha^{2}}.\label{freq}
\end{equation}
Then we can write equation (\ref{1/Bi}) as $1/B_{i,s}=n_{i}/\omega_{s}+\Delta_{s}$,
where $\Delta_s$ is the phase 

\begin{equation}
\Delta_{s}=-\frac{2s_{i}\mu_{\mathrm{B}}}{\mu}.\label{phase}
\end{equation}
From equations (\ref{freq}) and (\ref{phase}) we see that $\omega_{1}=\omega_{-1}$
and $\Delta_{1}=-\Delta_{-1}$. This means that the two sawtooth peaks
have the same frequency but different phase. The peaks amplitude are
obtained from equation (\ref{DeltaM}). Suppose the magnetic field is increased
so the last sorted position $f$ changes to $f-1$. For $\Delta n$
and $\Delta m_{P}$ we easily get $\Delta n=D/\mathcal{A}=2B/\phi$
and $\Delta m_{P}=D\mu_{\mathrm{B}}s_{f}/\mathcal{A}=2B\mu_{\mathrm{B}}s_{f}/\phi$,
so in general

\begin{equation}
A_{s}=\frac{-\mu+s\mu_{\mathrm{B}}B}{\phi}.\label{amp}
\end{equation}
We are now in position to express the two SO as an infinite series,
whose amplitude, frequency and phase are given by equations (\ref{freq}),
(\ref{phase}) and (\ref{amp}). We shall note $\omega\equiv\omega_{1}=\omega_{-1}=\mu^{2}/\alpha^{2}$
and $\Delta\equiv-\Delta_{1}=\Delta_{-1}=2\mu_{\mathrm{B}}/\mu$. Then we have

\begin{equation}
M_{SO}=\sum_{s=\pm1}A_{s}\sum_{p=1}^{\infty}\frac{1}{\pi p}\sin\left[2\pi p\omega\left(\frac{1}{B}+s\Delta\right)\right].\label{M sawtooth}
\end{equation}
Equation (\ref{M sawtooth}) gives the SO contribution to the MO. There
is still another oscillatory contribution, which comes from the continuous
oscillation in $\Omega_{C}$. This oscillation is not a SO, but an expression for it can be straightforward
obtained noticing that $M_{osc}=-\mathcal{A}^{-1}\left(\partial\Omega_{C}^{osc}/\partial B\right)_{\mu}$.
From equation (\ref{M sawtooth}) we get that $\Omega_{C}^{osc}$ should
be of the form $\Omega_{C}^{osc}=\sum_{s=\pm1}C_{s}\sum_{p=1}^{\infty}\left(\pi p\right)^{-2}\cos\left[2\pi p\omega\left(\frac{1}{B}+s\Delta\right)\right]$,
where $C_{s}$ satisfies $M_{osc}=-\mathcal{A}^{-1}\left(\partial\Omega_{C}^{osc}/\partial B\right)_{\mu}$,
so $2\omega C_{s}/\mathcal{A}B^{2}=-A_{s}$. Therefore the MO is given by

\begin{figure}
	\includegraphics[scale=0.32]{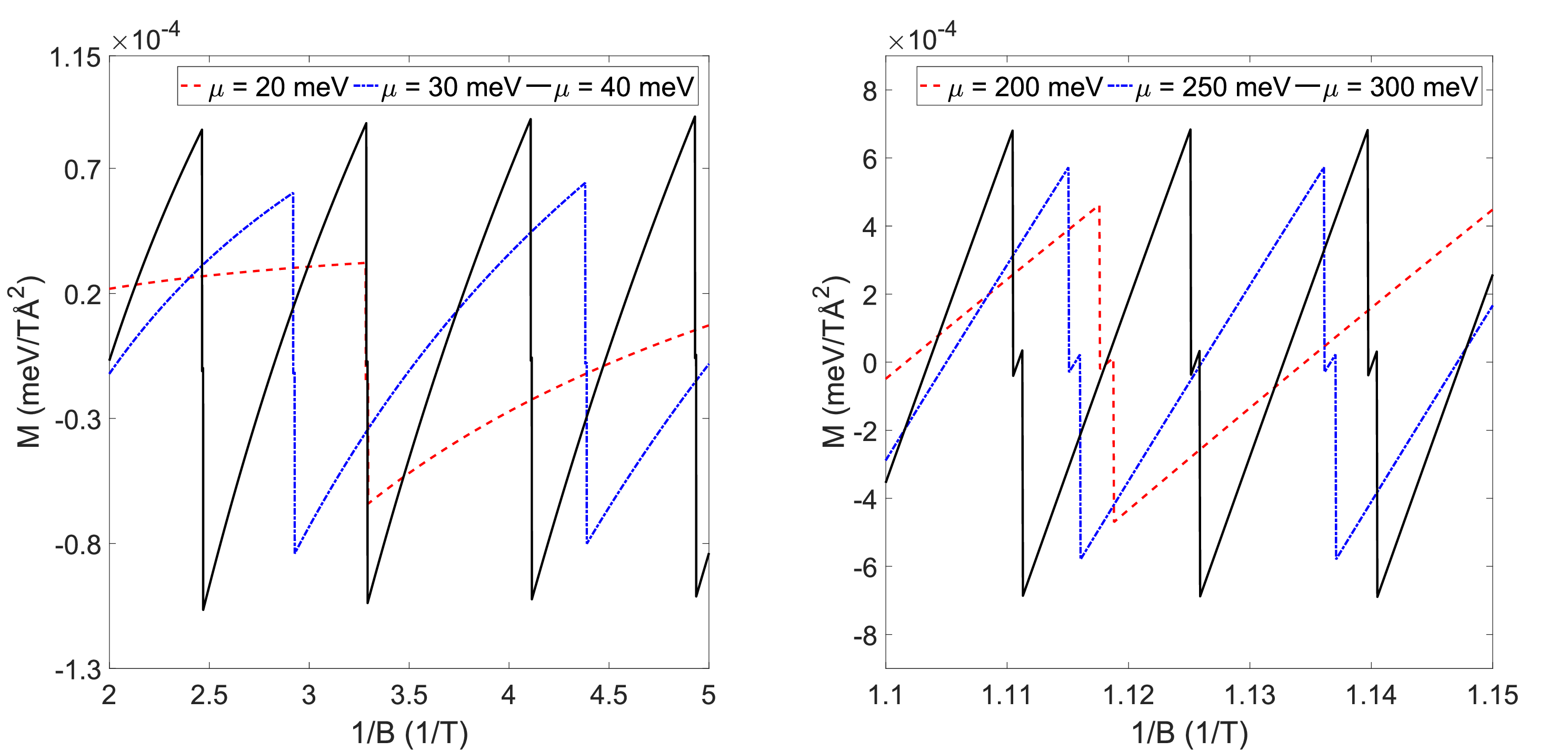}	
	\caption{\label{fig1}Magnetization given by equation (\ref{Mosc}), for (a) low $\mu$ and (b) high $\mu$, where $\mu$ is the Fermi energy.}	
\end{figure}

\begin{eqnarray}
\fl M_{osc} = \sum_{s=\pm1}\left\{ \left[\frac{B}{\omega\phi}\left(-\mu+\frac{3}{2}s\mu_{\mathrm{B}}B\right)\right]\sum_{p=1}^{\infty}\frac{1}{\left(\pi p\right)^{2}}\cos\left[2\pi p\omega\left(\frac{1}{B}+s\Delta\right)\right]\right.\nonumber\\
\left.+A_{s}\sum_{p=1}^{\infty}\frac{1}{\pi p}\sin\left[2\pi p\omega\left(\frac{1}{B}+s\Delta\right)\right]\right\}.\label{Mosc}
\end{eqnarray}
Equation (\ref{Mosc}) is in agreement with the result found in \cite{Sharapov2004},
where the oscillating part of the magnetization is expressed as an
infinite series. To obtain the total magnetization we also have to 
add the non-oscillatory contribution from both the valence and conduction
band. Nevertheless, it can be shown \cite{Sharapov2004} that in graphene, at $\mu>0$,
this contribution cancels and the total magnetization is just given
by equation (\ref{Mosc}). This result can be intuitively understood by noticing
that the non-oscillatory contribution to $M$ comes from the variation
of the degeneracy and energy as $B$ is changed (which gives a variation
proportional to $\sqrt{B}$, for $D\varepsilon\sim B^{3/2}$ and thus
$\delta D\varepsilon\sim B^{1/2}$). Ignoring the Zeeman splitting, the valence and conduction band energy levels are equal with opposite
sign, so their total contribution is canceled. Consequently, in order to
study the magnetization in graphene at $\mu>0$ it is sufficient to
work with equation (\ref{Mosc}). It is worth noting that in this formalism,
the spin splitting due to the Zeeman effect is already taken into account
in equation (\ref{Mosc}), so there is no need to introduce it as
a reduction factor. The spin splitting is usually neglected in the
MO calculations, but we will see that it can have a noticeable effect
in $M$ at high Fermi energy, with sufficient low temperature and/or high magnetic field.

\subsection{Spin splitting in the MO at zero temperature}

We shall now analyze equation (\ref{Mosc}) in more detail. First, notice
that each spin has a different amplitude $A_{s}$, although from equation
(\ref{amp}) we have $\left|A_{1}-A_{-1}\right|=\mu_{\mathrm{B}}B/\phi$, which
usually is very small. On the other hand, the MO are periodic in $1/B$
(as in the classical case), each spin peak with the same frequency
$\omega=\mu^{2}/\alpha^{2}$, but with a phase difference $2\Delta=4\mu_{\mathrm{B}}/\mu$.
These features can be seen in figure \ref{fig1}, where the MO given by equation (\ref{Mosc})
is plot for (a) low $\mu$ and (b) high $\mu$. We observe that, in order of magnitude, for
$\mu\sim10$ meV the MO are practically seen as one unique oscillation, while for $\mu\sim100$ meV
the spin splitting becomes noticeable. 
Moreover, we observe that at low $\mu$ there is a small curvature in the MO, which disappears at high $\mu$. This behavior can be explained by taking the ratio of amplitudes
of both series, given by $A_{s}^{cos}=B\left(-\mu+3s\mu_{\mathrm{B}}B/2\right)/\omega\phi$
and $A_{s}^{sin}=\left(-\mu+s\mu_{\mathrm{B}}B\right)/\phi$. Considering that
usually $B\mu_{\mathrm{B}}/\mu\ll1$, we get $A_{s}^{cos}/A_{s}^{sin}\simeq B/\omega$.
Thus, if $\mu>0.1$ eV we have $\omega=\mu^{2}/\alpha^{2}>10$ T and $A_{s}^{cos}/A_{s}^{sin}<0.1B\left[\unit{T}\right]$,
so unless $B$ is very high we have $A_{s}^{cos}/A_{s}^{sin}\ll1$. For smaller values of $\mu$ this would not be case, and for big $B$ one may even have $A_{s}^{cos}/A_{s}^{sin}>1$. This is seen in figure \ref{fig1}(a), where at low $\mu$ the cosine series in equation (\ref{Mosc}) produces the small curvature in the MO. Nevertheless, we will not take $\mu$ this low for then the Zeeman effect becomes unobservable. 
Therefore, given that we will be mainly interested with the spin splitting in the MO, which is seen at high $\mu$, we shall directly neglect the first term in equation (\ref{Mosc}). Then the MO reduce to equation (\ref{M sawtooth}) and the sine series can be evaluated to
obtain 

\begin{equation}
M=\sum_{s=\pm1}\frac{A_{s}}{\pi}\arctan\left\{ \cot\left[\pi\omega\left(\frac{1}{B}+s\Delta\right)\right]\right\} .\label{M0}
\end{equation}

\begin{figure}[h]
	\includegraphics[scale=0.35]{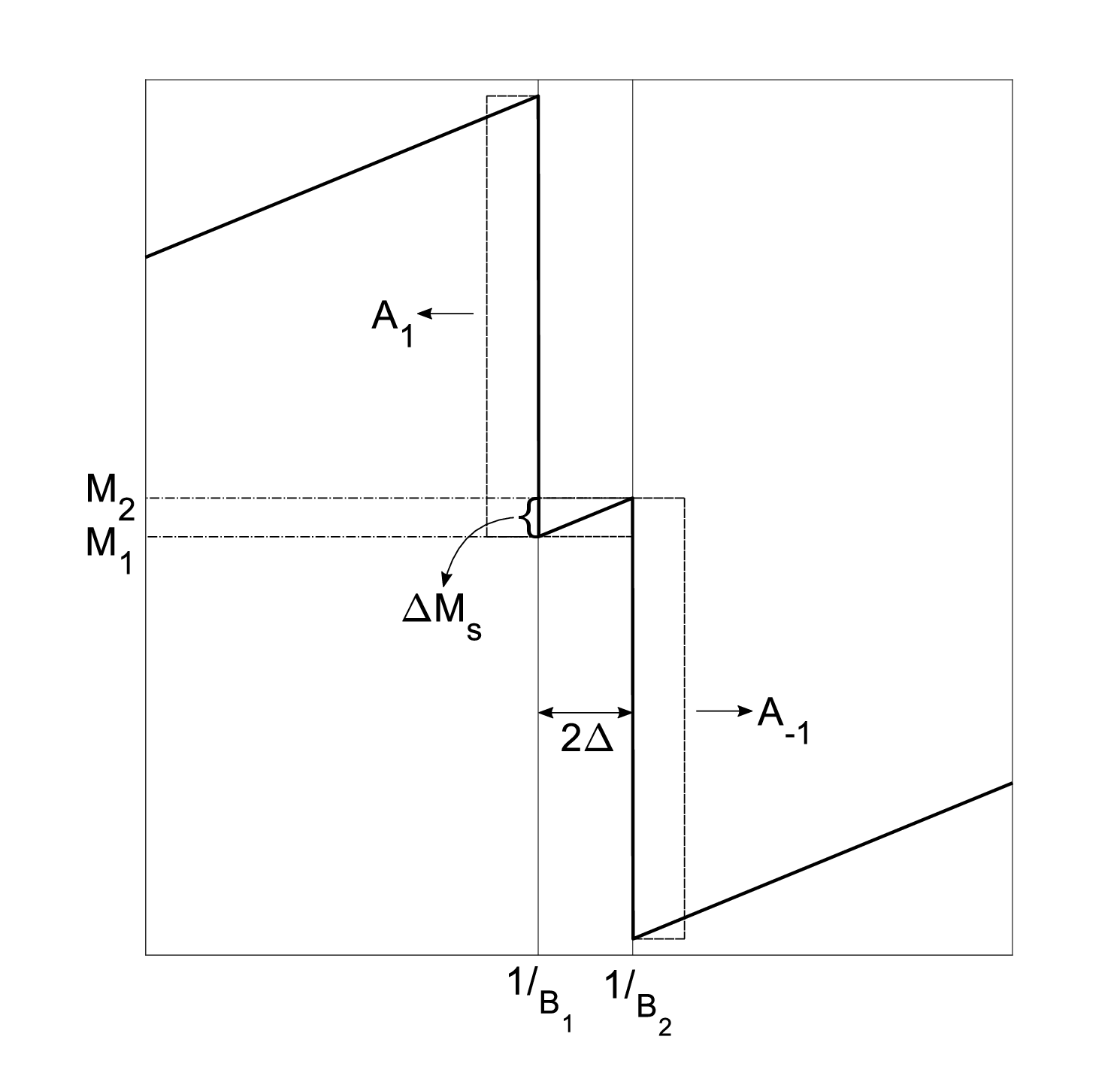}	
	\caption{\label{fig2}Scheme of two peaks due to the Zeeman effect, located at $1/B_1=l/\omega-\Delta$ and $1/B_2=l/\omega+\Delta$. We define the amplitude of splitting $\Delta M_s$ as shown, which measures the observation of the Zeeman effect in the MO. Notice also that the peaks have a phase difference $2\Delta=4\mu_{\mathrm{B}}/\mu$.}	
\end{figure}

We can calculate the amplitude of the spin splitting effect in the MO by taking into account
the low point of the first peak and the high point of the second peak,
as schematically indicated in figure \ref{fig2}. We consider the two peaks corresponding to a 
LL $l$, so that the first peak is located at $1/B_{1}=l/\omega-\Delta$,
while the second peak at $1/B_{2}=l/\omega+\Delta$. Then, from equation
(\ref{M0}) we get that at the low point of the first peak we have $M_{1}=A_{1}/2-A_{-1}\arctan\left[\cot\left(2\pi\omega\Delta\right)\right]/\pi$,
while at the high point of the second peak we have $M_{2}=A_{1}\arctan\left[\cot\left(2\pi\omega\Delta\right)\right]/\pi-A_{-1}/2$.
Thus the amplitude of splitting $\Delta M_{S}$ is given by

\begin{figure}[h]
	\includegraphics[scale=0.35]{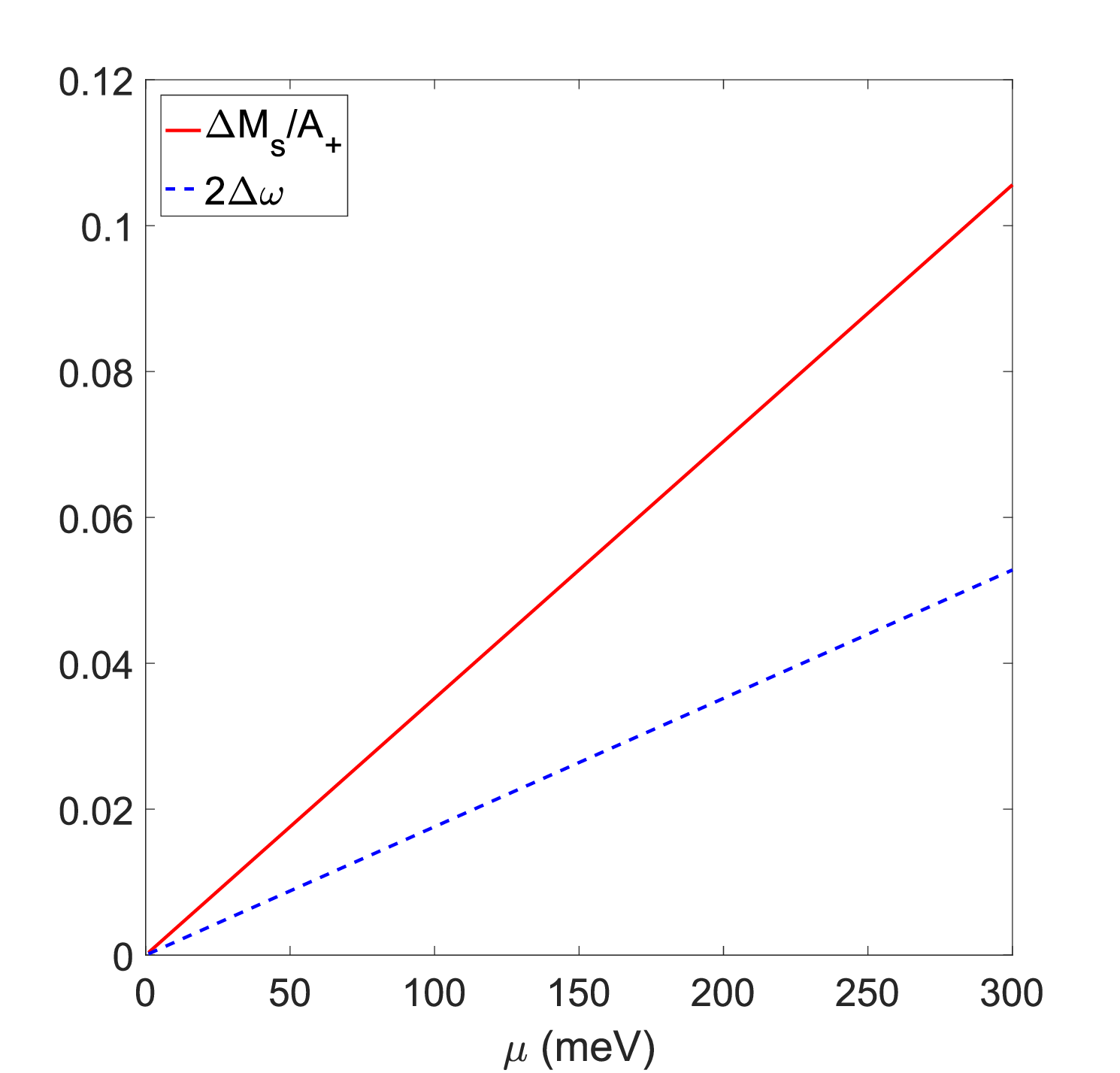}	
	\caption{\label{fig3}Plot of the parameters $\Delta M_{s}/A_{+}$ and $2\Delta\omega$
		as a function of the Fermi energy, where $\Delta M_{s}$ is given by equation (\ref{Delta Ms}) and $A_{+}\equiv\left|A_{1}+A_{-1}\right|/2=\mu/\phi$. The phase $\Delta$ and frequency $\omega$ are given by equations (\ref{freq}) and (\ref{phase}).}	
\end{figure}

\begin{equation}
\Delta M_{s}=A_{+}\left(1-\frac{2}{\pi}\arctan\left[\cot\left(2\pi\omega\Delta\right)\right]\right),\label{Delta Ms}
\end{equation}
where $A_{+}\equiv\left|A_{1}+A_{-1}\right|/2=\mu/\phi$. A better
measurement of the effect is to take the relation between $\Delta M_{s}$
and the amplitude of the peaks, which can be approximated as $A_{+}$
(because $A_1\simeq A_{-1}$). On the other hand,
we can also analyze the splitting peak width, which is given by the
phase difference $2\Delta=4\mu_{\mathrm{B}}/\mu$. This width also affect the
visualization of the spin splitting in $M$, for at low values of
Fermi energy, $2\Delta$ becomes insignificant. Then we can compare this
width with the oscillation period, given by $1/\omega=\alpha^{2}/\mu^{2}.$
In figure \ref{fig3} it is plotted $\Delta M_{s}/A_{+}$ and $2\Delta\omega$
as a function of the Fermi energy. We see that for $\mu\lesssim100$
meV we have $\Delta M_{s}/A_{+}\sim10^{-2}$, while $2\Delta\omega$
is also very small, which makes the spin splitting effect practically
unobservable. But when we get to higher Fermi energies, about $\mu\gtrsim250$
meV, we have $\Delta M_{s}/A_{+}\sim0.1$ so the effect starts to
become more evident. Consequently, in order to account for the Zeeman
effect in $M$ we will take Fermi energies above 200 meV. Notice also
that $\Delta M_{s}/A_{+}$ does not depend on $B$, so the effect
of the Zeeman splitting in the magnetization at zero temperature is
independent of the magnetic field. Nevertheless, we shall see that
the temperature makes the observation of the Zeeman splitting in the
magnetization be strongly dependent of $B$.

\section{MO at nonzero temperature}

The effects of nonzero temperature can be taken into account
in equation (\ref{M sawtooth}) by introducing the temperature reduction factor
$R_{T}$, which for these systems has the form \cite{Sharapov2004}

\begin{equation}
R_{T}=\frac{4\pi^{2}\mu pkT/\alpha^{2}B}{\sinh\left(4\pi^{2}\mu pkT/\alpha^{2}B\right)},\label{Rt}
\end{equation}
where $\alpha=\upsilon_{F}\sqrt{2e\hbar}$. Thus the MO at nonzero temperature becomes

\begin{equation}
M_{T}=\sum_{s=\pm1}A_{s}\sum_{p=1}^{\infty}\frac{R_{T}}{\pi p}\sin\left[2\pi p\omega\left(\frac{1}{B}+s\Delta\right)\right].\label{M temperature}
\end{equation}
In general, the series given by equation (\ref{M temperature}) cannot
be evaluated, but it can be simplified under some approximations.
The usual one is to take the limit $B\rightarrow0$, in which
case $1/\sinh\left(4\pi^{2}\mu pkT/\alpha^{2}B\right)\sim\exp\left(-4\pi^{2}\mu pkT/\alpha^{2}B\right)$.
Nevertheless, this limit implies low magnetic field, which in turn
makes it difficult to observe the MO. Instead we will impose the
low temperature limit, but such that the MO are still observable. 

To obtain
an approximation for $M_{T}$ at low temperatures, it is convenient
to start from the expression of the grand potential $\Omega$ at $T\neq0$
and $\mu>0$. It can be shown that in the absence of impurities, one can use the non-relativistic grand potential to obtain the magnetization \cite{Tabert2014}. Hence we start with

\begin{equation}
\Omega_{T}=-kT\int_{-\infty}^{\infty}\rho_{0}(\varepsilon)\ln\left[1+e^{\beta\left(\mu-\varepsilon\right)}\right]d\varepsilon,
\end{equation}
where $\beta=1/kT$ and $\rho_{0}(\varepsilon)=D\sum_{m}\delta\left(\varepsilon-\varepsilon_{m}\right)$
is the density of states (DOS) in the pristine case, where the summation is to be done considering the valence and conduction band. In this way we can separate
the contribution from both bands, so that $\Omega_{T}=\Omega_{V,T}+\Omega_{C,T}$,
where in general $\Omega_{T}=-kT\sum_{\varepsilon}D\ln\left[1+e^{\beta\left(\mu-\varepsilon\right)}\right]$.
Now, for $\mu>0$ and low temperature we always have $\beta\left(\mu-\varepsilon\right)\gg1$
for the valence band, so $\Omega_{V,T}\simeq\Omega_{V}\left(T=0\right)$
and $\partial\Omega_{V,T}/\partial B\simeq\partial\Omega_{V}\left(T=0\right)/\partial B$.
Hence the valence band magnetization is not affected by the temperature
under this conditions. Therefore, we shall omit it for the moment
and continue with the conduction band grand potential

\begin{equation}
\Omega_{C,T}=-kT\sum_{m=0}^{\infty}D\ln\left[1+e^{\beta\left(\mu-\varepsilon_{m}\right)}\right]=\sum_{m=0}^{\infty}\Omega_{T}^{m},
\end{equation}
where $\varepsilon_{m}=\alpha\sqrt{n_{m}B}-s_{m}\mu_{\mathrm{B}}B$ are the
conduction energy levels, and we defined $\Omega_{T}^{m}=-kTD\ln\left[1+e^{\beta\left(\mu-\varepsilon_{m}\right)}\right]$.
We have $\left(\partial\Omega_{T}^{m}/\partial B\right)_{\mu}=\Omega_{T}^{m}/B+D\left(\partial\varepsilon_{m}/\partial B\right)\left[1+e^{-\beta\left(\mu-\varepsilon_{m}\right)}\right]^{-1}$,
so the conduction magnetization $M_{C,T}=-\mathcal{A}^{-1}\left(\partial\Omega_{C,T}/\partial B\right)_{\mu}$
is given by

\begin{equation}
M_{C,T}=\frac{KTD}{\mathcal{A}B}\sum_{m=0}^{\infty}\ln\left[1+e^{\beta\left(\mu-\varepsilon_{m}\right)}\right]-\frac{D}{\mathcal{A}}\sum_{m=0}^{\infty}\frac{\partial\varepsilon_{m}}{\partial B}\frac{1}{1+e^{-\beta\left(\mu-\varepsilon_{m}\right)}}.\label{Mt}
\end{equation}
We now assume a Fermi energy $\varepsilon_{f}<\mu\leq\varepsilon_{f+1}$,
so the conduction magnetization at zero temperature is $M_{C}=-\mathcal{A}^{-1}\left(\partial\Omega_{C}/\partial B\right)_{\mu}=-\mathcal{A}^{-1}\sum_{m=0}^{f}\left[D\left(\varepsilon_{m}-\mu\right)/B+D\left(\partial\varepsilon_{m}/\partial B\right)\right]$. Thus
we can rewrite equation (\ref{Mt}) as

\begin{eqnarray}
\fl M_{C,T} =  M_{C}+\frac{KTD}{\mathcal{A}B}\sum_{m=0}^{f}\ln\left[1+e^{-\beta\left(\mu-\varepsilon_{m}\right)}\right]+\frac{D}{\mathcal{A}}\sum_{m=0}^{f}\frac{\partial\varepsilon_{m}}{\partial B}\frac{1}{1+e^{\beta\left(\mu-\varepsilon_{m}\right)}}\nonumber\label{Mt2}\\
 +\frac{KTD}{\mathcal{A}B}\sum_{m=f+1}^{\infty}\ln\left[1+e^{\beta\left(\mu-\varepsilon_{m}\right)}\right]-\frac{D}{\mathcal{A}}\sum_{m=f+1}^{\infty}\frac{\partial\varepsilon_{m}}{\partial B}\frac{1}{1+e^{-\beta\left(\mu-\varepsilon_{m}\right)}}.
\end{eqnarray}
We can simplify equation (\ref{Mt2}) by noticing that the term with
the logarithm is in general very small. Indeed, we always have $\ln\left[1+e^{-\beta\left(\mu-\varepsilon_{m}\right)}\right]\leq\ln2<1$
if $m\leq f$ and likewise $\ln\left[1+e^{\beta\left(\mu-\varepsilon_{m}\right)}\right]<\ln2$
if $m>f$. Furthermore, we have $kTD/\mathcal{A}B=2kT/\phi=2kTe/h\sim10^{-7}T\left[\unit{K}\right]\;\textrm{meV/T}{\textrm{\AA}}^2$,
which for low $T$ is much smaller than the magnetization at zero
temperature (see figure \ref{fig1}). On the other hand, at low temperatures each exponential term in equation (\ref{Mt2})
is in general very small unless $\mu$ is close to $\varepsilon_{m}$.
Thus we can expand each $\varepsilon_{m}$ around $B_{m}$, where $B^{-1}_m=n_m/\omega+s_m\Delta$ and
$\varepsilon_{m}\left(B_{m}\right)=\mu$, so $\left(\mu-\varepsilon_{m}\right)\simeq-\mu\left(B-B_{m}\right)/2B_{m}$.
Moreover, $\left(\partial\varepsilon_{m}/\partial B\right)\simeq\left[\mu-s_{m}\mu_{\mathrm{B}}B\right]/2B$,
so from equation (\ref{amp}) we get $\mathcal{A}^{-1}\left(\partial\varepsilon_{m}/\partial B\right)\simeq-A_{s_{m}}/D.$
Finally, if we take into account the valence contribution
to the magnetization (which we show it is not altered by the temperature),
we have the total magnetization $M_{T}=M_{V,T}+M_{C,T}=M_{V}+M_{C,T}$.
Thus from equation (\ref{Mt2}) we get

\begin{figure}[t]
	\includegraphics[scale=0.35]{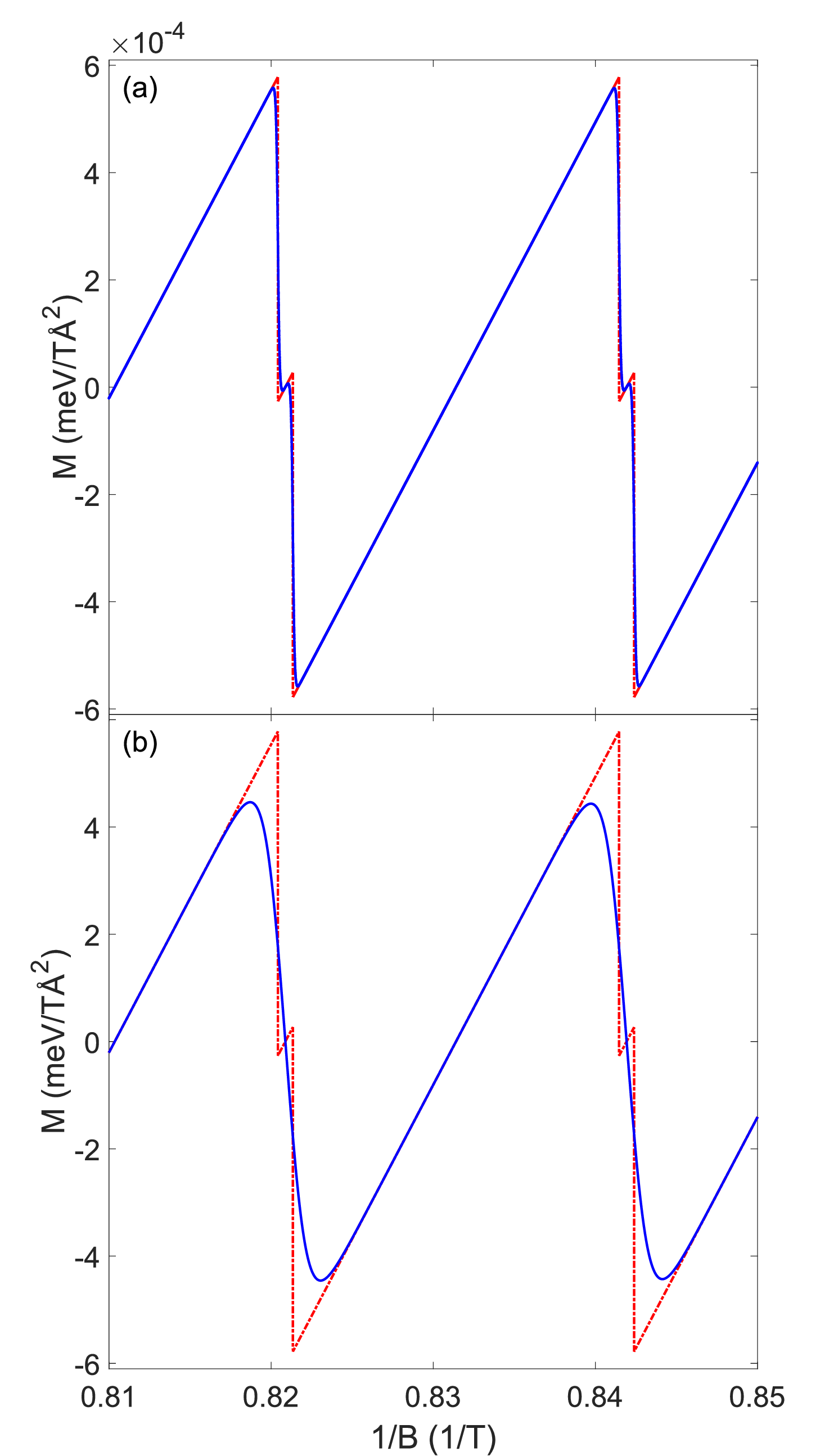}	
	\caption{\label{fig4}Magnetization given by equation (\ref{MT}), for $\mu=250$ meV. The red dashed line corresponds to the zero temperature case, while the solid blue line corresponds to (a) $T=0.1$ K and (b) $T=1$ K.}	
\end{figure}

\begin{equation}
M_{T} \simeq M-\sum_{m=0}^{f}\frac{A_{s_{m}}}{1+e^{-\beta\mu\left(B-B_{m}\right)/2B_{m}}}+\sum_{m=f+1}^{\infty}\frac{A_{s_{m}}}{1+e^{\beta\mu\left(B-B_{m}\right)/2B_{m}}}.\label{Mt aprox}
\end{equation}
where $M=M_{V}+M_{C}$ is the net magnetization at zero temperature,
given by equation (\ref{M0}). Equation (\ref{Mt aprox}) corresponds to the case $\varepsilon_{f}<\mu\leq\varepsilon_{f+1}$,
which implies $B_{f+1}\leq B<B_{f}$. Thus the low temperature effect is to
introduce a factor proportional to $\left[1+e^{-\beta\mu\left(B-B_{m}\right)/2B_{m}}\right]^{-1}$
if $B<B_{m}$, or proportional to $\left[1+e^{\beta\mu\left(B-B_{m}\right)/2B_{m}}\right]^{-1}$
if $B>B_{m}$. In this way we can generalize equation (\ref{Mt aprox}) to all $B$ and get the result (see the appendix for details)

\begin{equation}
\fl M_{T}=\sum_{s}\frac{A_{s}}{\pi}\arctan\left\{ \cot\left[\pi\omega\left(\frac{1}{B}+s\Delta\right)+\sum_{n}\frac{\pi}{1+e^{-\beta\mu\left(B-B_{n,s}\right)/2B_{n,s}}}\right]\right\},\label{MT}
\end{equation}
where $B_{n,s}^{-1}=n/\omega-s\Delta$. In the practice it is sufficient
to take only the $B_{n}$ between the range of magnetic fields considered.

In figure \ref{fig4} it is show the magnetization given by equation (\ref{MT}) for $\mu=250$ meV and
temperatures (a) $T=0.1$ K, (b) $T=1$ K.
We can see that as the temperature increases, the magnetization
at $T\neq0$ broadens, similar to the Fermi-Dirac
distribution with the DOS. Of course,
this is exactly what is expected from the result given by equation (\ref{MT}).
Moreover, in figure \ref{fig4}(a) we see that at very low temperature $T=0.1$ K one
can still appreciate the effect of the spin splitting in the magnetization,
which results in a small bump around the two peaks. Nevertheless,
as the temperature is increased, this bump disappears and the MO behave as
maxima and minima around the center of the two peaks at zero temperature.
Thus in this regime there would be no noticeable effect of the spin
splitting in the magnetization. This is strongly dependent
not only on the temperature, but also on the magnetic field and Fermi
energy. In other words, for high magnetic fields, the temperature at which the spin splitting is negligible increases. It is also worth noting that the introduction of disorder in the system increases the damping of the MO, which would make it even harder to observe the spin splitting.

\subsection{Spin splitting in the MO at nonzero temperature}

An estimation for the temperature at which the spin splitting would
not be observable in the MO can be obtained from equation (\ref{MT}). We shall define
$x\equiv1/B$, and consider two peaks at $x_{1}$ and $x_{2}$, corresponding
to the same LL, with different spin, so in general $x_{1}=l/\omega-\Delta$
for the spin up, and $x_{2}=l/\omega+\Delta$ for the spin down (see
equation (\ref{1/Bi})). Then, between the two peaks, only the exponential
terms corresponding to $x_{1}$ and $x_{2}$ would be appreciable,
so for $x_{1}<x<x_{2}$ equation (\ref{MT}) becomes

\begin{equation}
M_{T}=M-\frac{A_{1}}{1+e^{\beta\mu\left(x-x_{1}\right)/2x}}+\frac{A_{-1}}{1+e^{-\beta\mu\left(x-x_{2}\right)/2x}},\label{Mt spin}
\end{equation}
where $M$ is given by equation (\ref{M0}), and we have rewrite $\left(B-B_{m}\right)/2B_{m}=-\left(x-x_{m}\right)/2x$.

\begin{figure}[t]
	\includegraphics[scale=0.3]{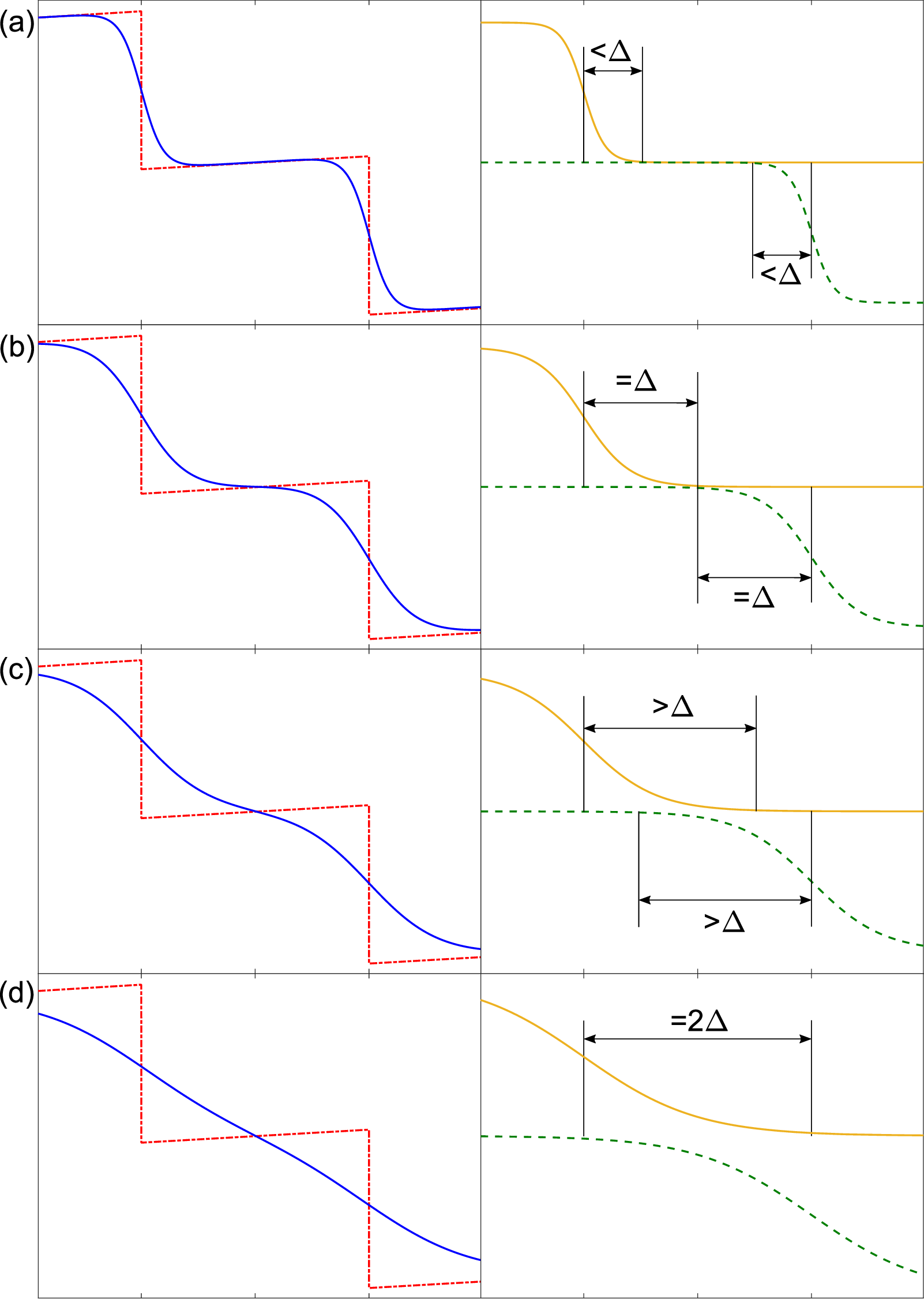}	
	\caption{\label{fig5}Spin splitting effect in the MO. In the left it is the magnetization at zero temperature (red dashed line) and at different temperatures (blue solid line). In the right it is plotted the first (orange solid line) and second (green dashed line) exponential in equation (\ref{Mt spin}). The four cases correspond to the exponentials width being (a) $<\Delta$, (b) $=\Delta$, (c) $>\Delta$, (d) $=2\Delta$, where $2\Delta$ is the spin peaks separation.}	
\end{figure}
From equation (\ref{Mt spin}) we can see that the temperature effect behaves
like a Fermi-Dirac distribution. The difference is that in this case
our parameter is $x=1/B$, instead of the energy, and each exponential
is center around the position of the peak $x_{m}$. Thus we
can write $\beta\mu\left(x-x_{m}\right)/2x=\tilde{\beta}\left(x-x_{m}\right)$,
where $\tilde{\beta}=1/k\tilde{T}\equiv\beta\mu/2x$
can be considered the corresponding \emph{temperature} parameter,
with defines the broadening of each exponential term in equation (\ref{Mt spin}).
This shows that the broadening depends not only on the real temperature
$T$, but also on the Fermi energy $\mu$ and the magnetic field $x=1/B$.
In this way we can get an estimation to how the temperature affects
the observation of the spin splitting in the magnetization, which
is dictated by how the width of the exponentials relates to the
spin peaks separation $x_{2}-x_{1}=2\Delta$. This can be seen in
figure \ref{fig5}, where for different temperatures we plotted the magnetization
(left) and the exponentials that appear in equation (\ref{Mt spin}) (right).
Then we can identify four situations. In the first case,
figure \ref{fig5}(a), the width is $<\Delta$ and one can clearly
appreciate the spin splitting in the magnetization at $T\neq0$. The
second case, figure \ref{fig5}(b), corresponds to the case when the width is
$\Delta$, so the exponentials began to overlap. Then, for a width
$>\Delta$ (but $<2\Delta$) in figure \ref{fig5}(c), the spin splitting effect
starts to disappear, although one could still notice a small change
of curvature in $M_{T}$. Finally, when the width is $\geq2\Delta$ as
in figure \ref{fig5}(d), the spin splitting becomes unobservable in $M_{T},$
and one is left with what appears as one unique oscillation around
the center $\left(x_{1}+x_{2}\right)/2$. 

Now, for a Fermi-Dirac distribution
of the form $\left[1+e^{(y-y_{0})/a}\right]^{-1}$, the width $w$ around
$y_{0}$ is determined by the condition $1/\left[1+e^{w/a}\right]=\sigma$, where $\sigma\ll1$ is the appreciation considered. From our experience, it is sufficient to take $\sigma\sim10^{-2}$, which implies a width $w\sim5a$. Therefore, following figure \ref{fig5}(d), the
critical temperature for which the spin splitting becomes unobservable
satisfies $2\Delta\simeq5k\tilde{T}$.
Given that $k\tilde{T}=2kTx/\mu$, and $2\Delta=4\mu_{\mathrm{B}}/\mu$, we
get the temperature $T_{s}=2\mu_{\mathrm{B}}/5kx$. This result basically
means that, in order of magnitude, the thermal energy $kT$ equals
the spin splitting energy $2\mu_{\mathrm{B}}/x$. One should replace in $x$
the value at which the broadening is computed. If the peak at $x_{1}$ is considered, the corresponding exponential factor is $\tilde{\beta}\left(x-x_{1}\right)$,
and the condition $2\Delta\simeq5k\tilde{T}$ implies to take
$x=x_{2}$. On the other hand, if we take the peak $x_{2}$, we should take $x=x_{1}$.
This gives two different temperatures $T_{s,1}\simeq2\mu_{\mathrm{B}}/5 Kx_{2}$
and $T_{s,2}\simeq2\mu_{\mathrm{B}}/5 Kx_{1}$, but $\left|T_{s,1}-T_{s,2}\right|=T_{s,1}\left(2\Delta/x_{1}\right)=T_{s,2}\left(2\Delta/x_{2}\right)$
with $\Delta/x_{i}\ll1$ for both cases. Thus in general $T_{s,1}\simeq T_{s,2}$,
and we can take the average temperature $T_{s}=\left(T_{s,1}+T_{s,2}\right)/2$,
so

\begin{equation}
T_{s}\simeq\frac{2}{5}\frac{\mu_{\mathrm{B}}B}{k},\label{Ts}
\end{equation}
where $B=\omega/l$ is the center of the spin peaks. Equation (\ref{Ts})
gives the maximum temperature to observe the spin splitting in the
MO. Given that $B=\omega/l=\mu^{2}/l\alpha^{2}$, then $T_{s}$
is proportional to $\mu^{2}$, so high Fermi energy favors the observation
of the spin splitting in the magnetization. It should be noted that the condition
given by equation (\ref{Ts}) has to be taken into account alongside with
the need of high $\mu$ (about $\mu\gtrsim100$ meV, see figure \ref{fig1}) to observe the spin splitting at
zero temperature. Therefore, for practical terms, equation (\ref{Ts})
should only be applied when $\mu\gtrsim100$ meV. 

\begin{figure}[t]
	\includegraphics[scale=0.35]{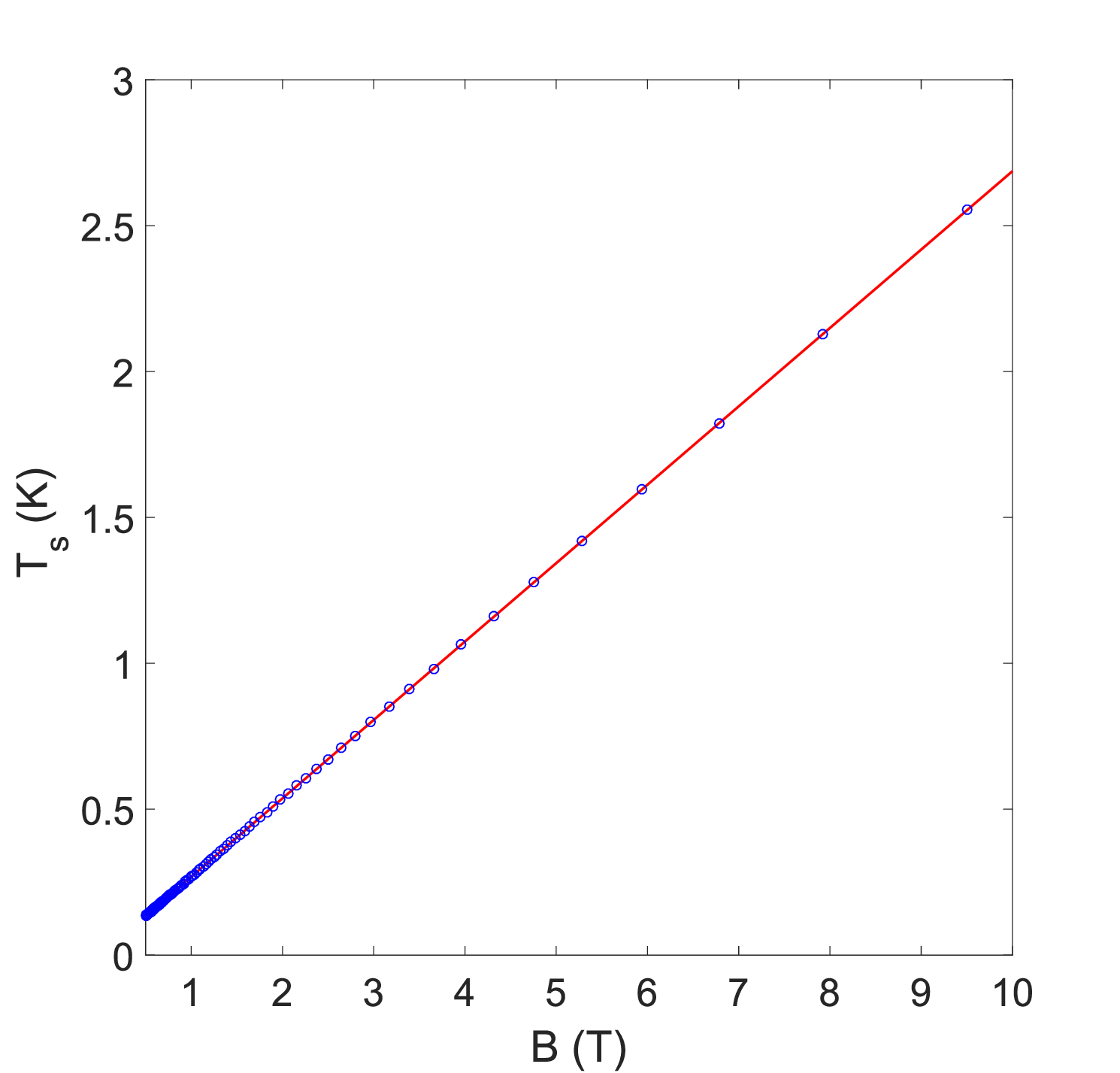}	
	\caption{\label{fig6}Spin temperature $T_s$, given by equation (\ref{Ts}), as a function of the magnetic field $B$ (solid line) and for $\mu=250$ meV (scatter), where $B=\omega/l=\mu^2/l\alpha^2$ at the center of the spin peaks corresponding to the LL $l$.}	
\end{figure}

In figure \ref{fig6} we plotted equation (\ref{Ts}) as a function
of the magnetic field $B$ (solid line) and for $\mu=250$ meV (scatter), where $B=\omega/l=\mu^2/l\alpha^2$. We see that, in general, $T_{s}$
is very low unless $B$ is very high. For instance, at $\mu=250$ meV, we have that $B\sim1$ T gives $T_{s}\sim0.3$ K. Thus, even for high magnetic field, it
requires a very low temperature to observe the spin splitting in the MO. Furthermore, if one also considers disorder in graphene, one should expect that the spin temperature becomes lower with increasing disorder.

\subsection{MO extrema shift}

Once the temperature is such that $T>T_{s}$ in equation (\ref{Ts}),
the spin splitting becomes unobservable. Then the MO behave as if
the spin splitting is neglected, with the oscillation been around $B=\omega/l$. This is just the usual behavior when the MO
are studied in graphene without considering the Zeeman effect. Thus, because we shall take $T>T_{s}$ from now on, we will neglect the spin splitting.  Then, what
one observe is, essentially, that the MO are broadened and reduced
as the temperature increases. This reduction depends on the temperature,
as well as the magnetic field and Fermi energy. The higher the magnetic
field, the more temperature it takes to reduce the
MO, and vice versa. From this pattern we can analyze different parameters in the MO, in which case equation (\ref{MT})
will prove useful. 

First of all, notice that as the MO are broadened,
the maxima and minima of the oscillations shift from $x_{n}\equiv B_{n}^{-1}=n/\omega$
(see equation (\ref{1/Bi}), neglecting the spin splitting) to $x_{n}-p\delta$, where $p=1(-1)$ for the maxima (minima) and $\delta=\delta\left(T,B,\mu\right)$
is the shift parameter which in general depends on the three variables
$T,\,B$ and $\mu$. For a fixed $\mu$, we can obtain the function
$\delta$ by the extrema condition $\partial M_{T}/\partial x=0$, where $x=1/B$.
Now, neglecting the spin splitting, equation (\ref{MT}) becomes

\begin{equation}
M_{T}=-\frac{2\mu}{\pi\phi}\arctan\left[\cot\left(\pi\omega x+\sum_{n}\pi f_{n}\right)\right],\label{MTSP}
\end{equation}
where we defined  $f_{n}\equiv\left[1+e^{\beta\mu\left(x-x_{n}\right)/2x}\right]^{-1}$. Then, the extrema condition $\partial M_{T}/\partial x=0$ can be
written as

\begin{equation}
1=\frac{\beta\mu}{8\omega x^{2}}\sum_{n}x_{n}\textrm{sech}^{2}\left[\frac{\beta\mu\left(x-x_{n}\right)}{4x}\right].\label{x extreme}
\end{equation}
The $x=1/B$ that
satisfy equation (\ref{x extreme}) give all the new maxima and minima
in the MO at $T\neq0$. Notice that at zero temperature $\beta\rightarrow\infty$
and equation (\ref{x extreme}) implies $x\rightarrow x_{n}$, as expected.
In general, for a peak at $x_{l}=l/\omega$, to obtain its shift it is sufficient
to consider only the $n$ close to $l$ (say $l-2<n<l+2$) in the
summation of equation (\ref{x extreme}). The new extrema will be located
at $x=x_{l}-p\delta_{l}$, where $\delta_{l}$ is the corresponding
shift and $p=1(-1)$ for the maxima (minima). In this way we can rewrite equation (\ref{x extreme}) as a function
of $\delta_{l}$ and the considered peak $x_{l}$: 

\begin{equation}
1 =\frac{\beta\mu}{8\omega\left(x_{l}-p\delta_{l}\right)^{2}}
 \sum_{n}x_{n}\textrm{sech}^{2}\left[\frac{\beta\mu\left(x_{l}-x_{n}-p\delta_{l}\right)}{4\left(x_{l}-p\delta_{l}\right)}\right].\label{delta extreme}
\end{equation}
\begin{figure}[t]
	\includegraphics[scale=0.32]{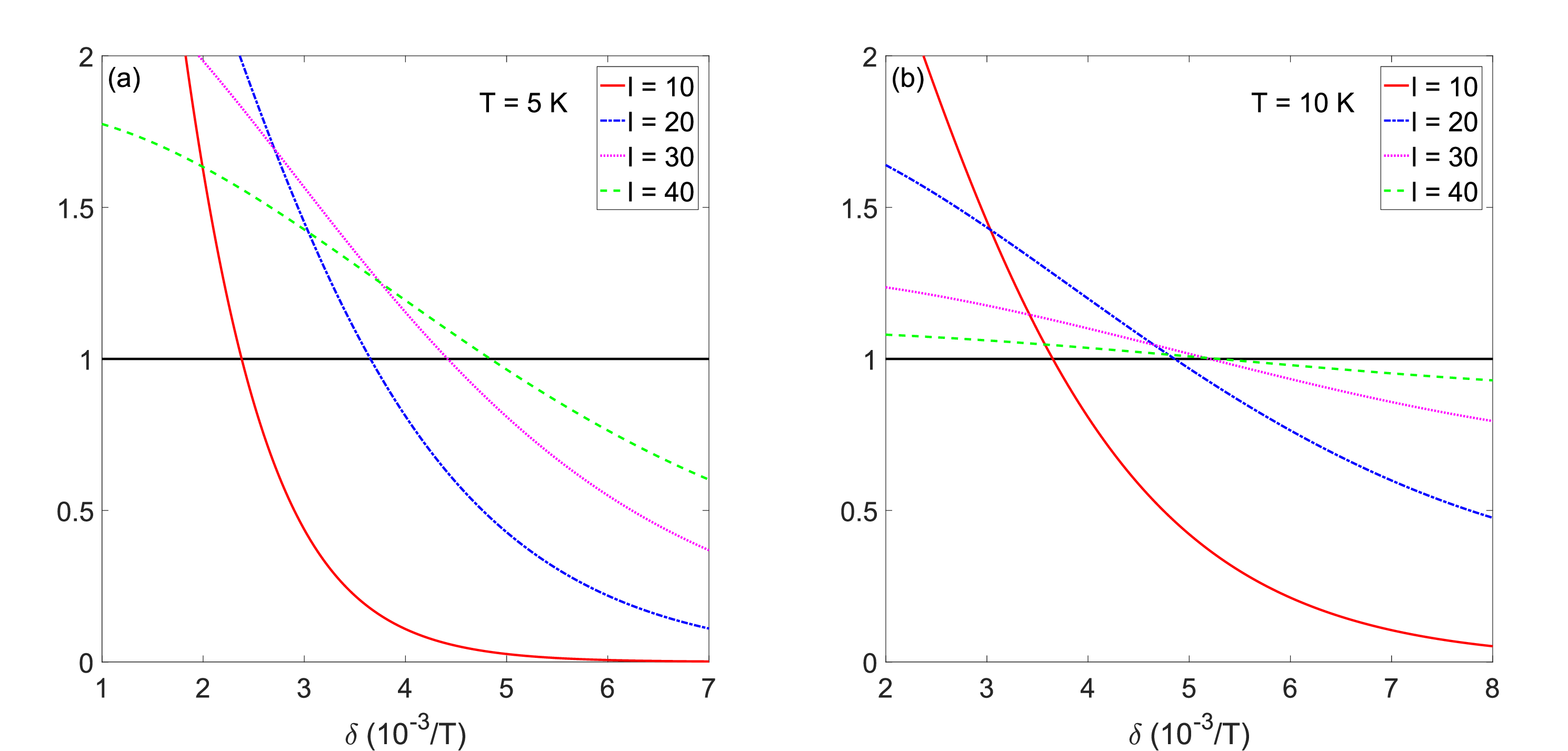}	
	\caption{\label{fig7}Graphical solution of equation (\ref{delta extreme}) for different LL $l$ and temperatures (a) $T=5$ K, (b) $T=10$ K. All cases correspond to $\mu=250$ meV and maxima shift, so $p=1$ in equation (\ref{delta extreme}).}	
\end{figure}
In figure \ref{fig7} it is show the graphical solution of equation (\ref{delta extreme})
for different values of temperature and $l$, for $\mu=250$ meV and
$p=1$ (maxima shift).

As we can see, the shift $\delta$ clearly depends on the LL $l$
and thus on the magnetic field. It increases with the temperature
and decreases with the magnetic field. This means that for non zero temperature,
the MO are not anymore periodic in $1/B$. Indeed, $\delta$ as defined
is always measured from the position of the peaks at zero temperature, whose distance
between one another is always the same and identical to $1/\omega$ (the
period). Consequently, if at $T\neq0$ the shift $\delta_{l}$
is not equal for all $l$, then the extrema separation will not be constant
and thus not periodic as a function of $1/B$. Nevertheless, it should
be notice that this effect is very small, because for close $l_{1}$
and $l_{2}$ one usually has $\left|\delta_{1}-\delta_{2}\right|\ll1/\omega$. On the other hand,
in figure \ref{fig7}(b) we see that for the cases $l=30$ and
$l=40$, the shift $\delta$ reaches an steady value of about $\delta\sim5.3\times10^{-3}\unit{T^{-1}}$. This
means that the extrema in the MO will tend to be located between the peaks at zero temperature and the zeros in the magnetization. Indeed, the zeros
in the MO occur at $B_{0n}^{-1}=\left(2n+1\right)/2\omega$. Thus, the distance
between the peaks and the zeros is $\left(B_{n}^{-1}-B_{0n}^{-1}\right)=1/2\omega$,
which is about $\sim1.05\times10^{-2}\unit{T^{-1}}$ for $\mu=250$ meV (as considered
in figure \ref{fig7}), so the midpoint distance is about $\sim5.3\times10^{-3}\unit{T^{-1}}$. 

\begin{figure}[t]
	\includegraphics[scale=0.35]{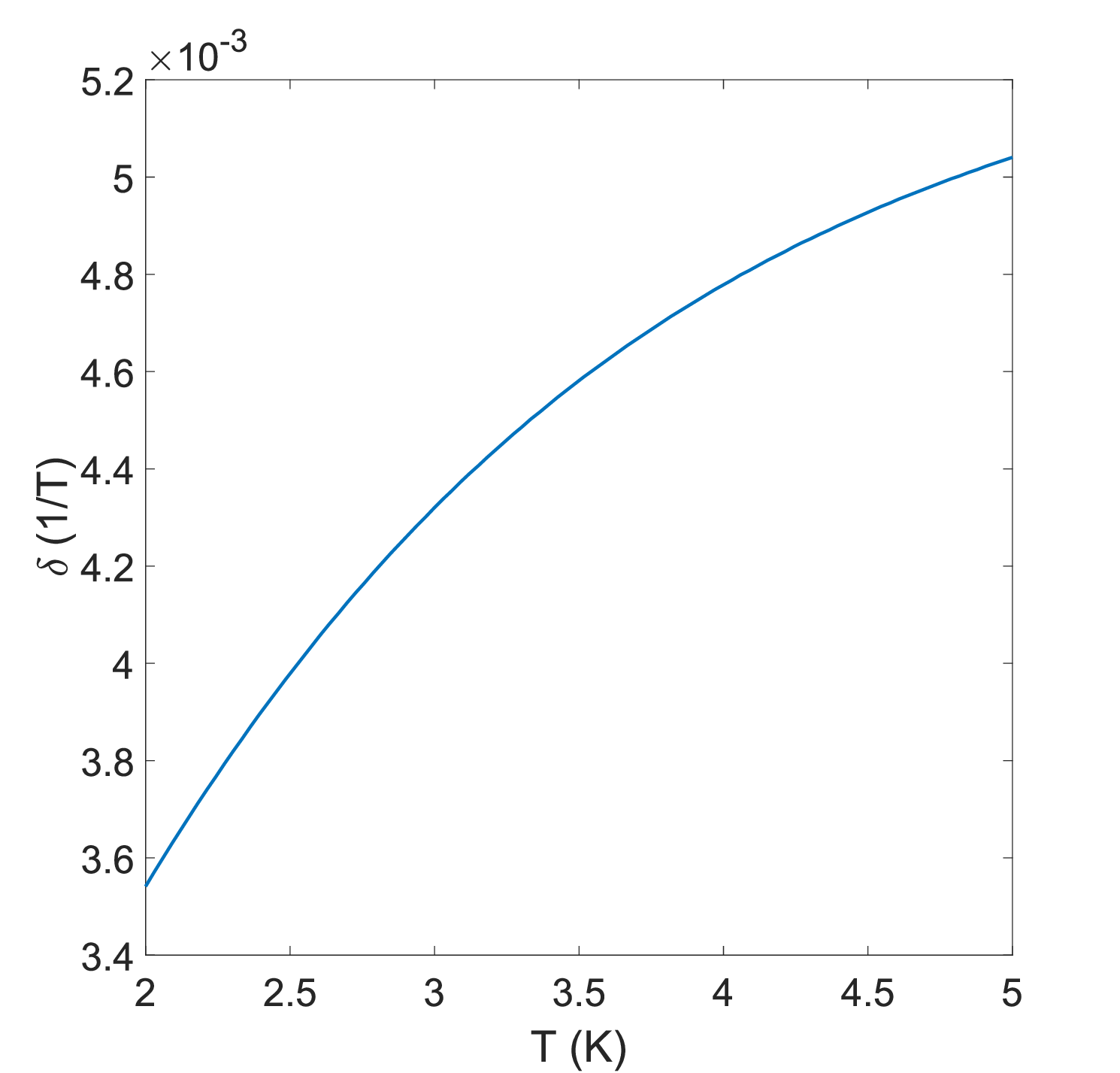}	
	\caption{\label{fig8} Numerical solution of equation (\ref{delta extreme}) for $\delta_l$ as a function of the temperature, where $\mu=250$ meV and $B=1/x_l=\omega/l=1.01$ T (corresponding to LL $l=47$).} 	
\end{figure}

We can also analyze the shift as a function of the temperature, for a particular LL. This can be seen in figure \ref{fig8}, which shows the numerical solution of equation (\ref{delta extreme}), for $\mu=250$ meV and $B=1/x_l=\omega/l=1.01$ T (corresponding to LL $l=47$). As expected, we see that a low temperature $\delta_l$ tends to zero, and increases with $T$. This behavior could be useful to measure temperature changes from the MO. 

\subsection{MO envelope}

If, at a given temperature, $\delta$ is know as a function of $B$,
then the envelope of the MO can be easily obtained. We
start considering $x_{1}<x<x_{2},$ where $x_{1}$ and
$x_{2}$ are two adjacent peaks at zero temperature, so in general
$x_{1}=\left(l-1\right)/\omega$ and $x_{2}=l/\omega$.
Then, from equation (\ref{MTSP}) we can write

\begin{equation}
M_{T} \simeq M+\frac{2\mu}{\phi}\left(\sum_{n=0}^{l-1}\frac{1}{1+e^{\beta\mu\left(x-n/\omega\right)/2x}}\right.
\left.-\sum_{n=l}^{\infty}\frac{1}{1+e^{-\beta\mu\left(x-n/\omega\right)/2x}}\right),\label{Mt env}
\end{equation}
where $M$ is given by equation (\ref{M0}), which without spin splitting is $M=-2\mu\arctan\left[\cot\left(\pi\omega x\right)\right]/\pi\phi$. To obtain the positive
envelope corresponding to the maxima in the MO, we just
have to eliminate the oscillatory part in (\ref{Mt env}). For $M$, this
can be done by replacing $x=l/\omega-\delta$,
the position at which the maximum occurs, where $\delta$ is the shift
obtained from equation (\ref{delta extreme}). Thus from equation (\ref{M0})
we get 
\begin{equation}
M_{\delta}=\frac{2\text{\ensuremath{\mu}}}{\pi\phi}\arctan\left[\cot\left(\pi\omega\delta\right)\right]\label{M delta},
\end{equation}
where we defined $M_{\delta}\equiv M\left(l/\omega-\delta\right)$.
Notice that $M_{\delta}$ is independent of $l$. On the
other hand, for the exponentials in equation (\ref{Mt env}), we need
to replace $x=l/\omega-\delta$ only in the numerator.
Thus, defining $m=l-n$ we get

\begin{eqnarray}
\fl \left[\sum_{m=1}^{l}\left(\frac{1}{1+e^{\beta\mu\left(m/\omega-\delta\right)/2x}}-\frac{1}{1+e^{\beta\mu\left(m/\omega+\delta\right)/2x}}\right)-\frac{1}{1+e^{\beta\mu\delta/2x}}\right.\nonumber\\
-\left.\sum_{m=l+1}^{\infty}\frac{1}{1+e^{\beta\mu\left(m/\omega+\delta\right)/2x}}\right].\label{exp up}
\end{eqnarray}
The last term in equation (\ref{exp up}) can be neglected because it
is very small if $x_{1}<x<x_{2}$. In this way, regrouping the exponentials we get the following expression
for the positive envelope

\begin{figure}
	\includegraphics[scale=0.35]{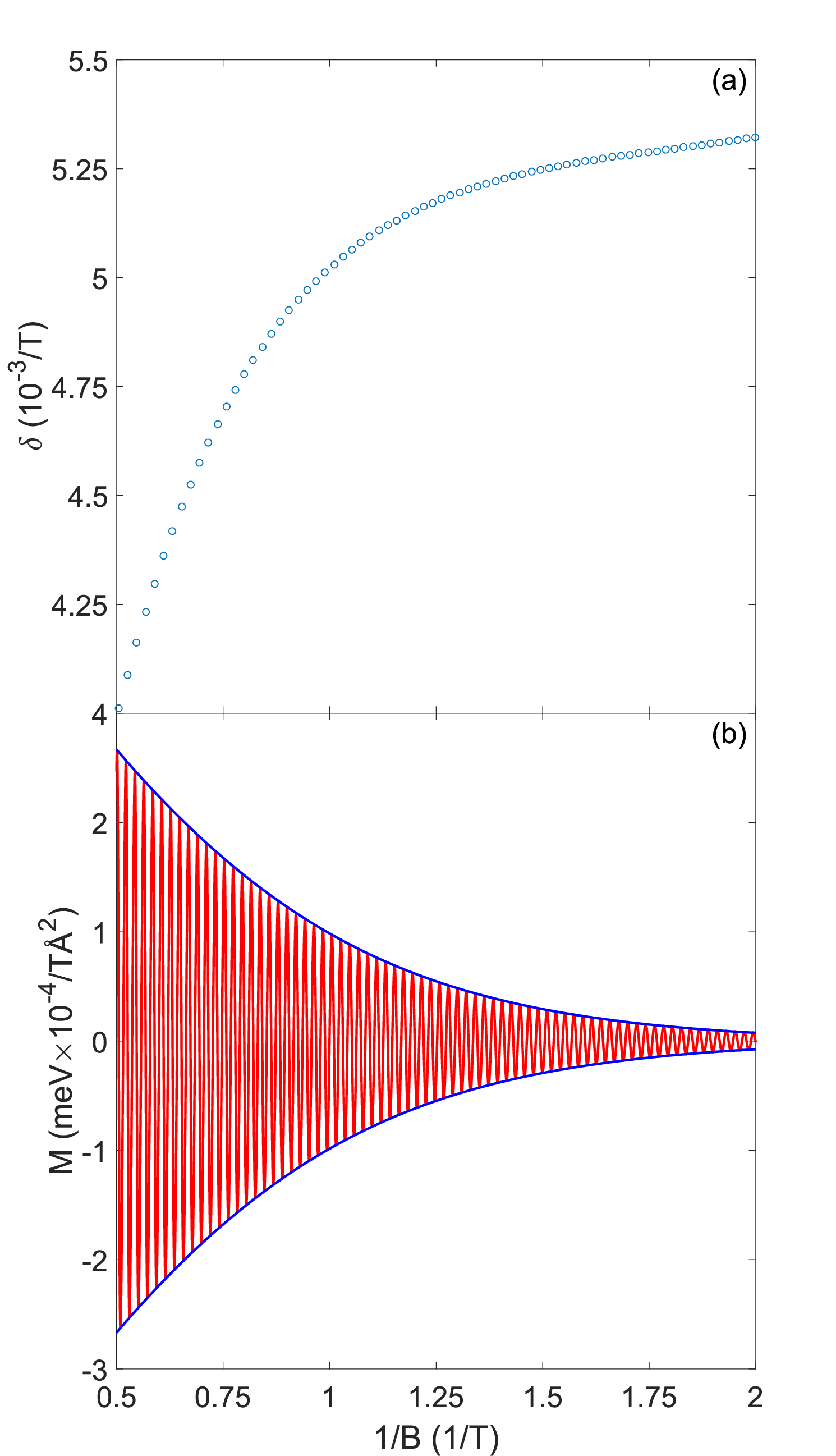}	
	\caption{\label{fig9} For $T=5$ K and $\mu=250$ meV: (a) Numerical solution of equation (\ref{delta extreme}), where $\delta\left(x\right)$ is obtained 
		by replacing $x=l/\omega$. The points were fitted with a Boltzmann function. (b) MO with the positive $E_{+}$ and negative $E_{-}=-E_{+}$ envelope, where $E_{+}$ is given by equation (\ref{E+}). The shift function $\delta(B)$ used was obtained from the fitting of (a).} 	
\end{figure}

\begin{eqnarray}
\fl E_{+} =\frac{2\mu}{\pi\phi}\left\{ \arctan\left[\cot\left(\pi\omega\delta\right)\right]-\frac{\pi}{1+e^{\beta\mu\delta/2x}}\right.\nonumber\\
\left.+\sum_{m\geq1}\frac{\pi\sinh\left[\beta\mu\delta/2x\right]}{\cosh\left[\beta\mu\delta/2x\right]+\cosh\left[\beta\mu m/2\omega x\right]}\right\}.\label{E+} 
\end{eqnarray}
It should be
notice that although this envelope was originally obtained for $x_{1}<x<x_{2}$,
the expression given by equation (\ref{E+}) is directly generalized and
valid for all magnetic fields. The negative envelope is just $E_{-}=-E_{+}.$
Although there is a summation involved in $E_{+},$ in the practice
it is sufficient to consider only the first few terms. In fact, the terms $m>1$ give corrections
only if we are at high temperature and/or low magnetic field, at which
the MO are small. 

The envelope given by equation (\ref{E+}) depends on the shift
$\delta$, which in general depends on the magnetic field, as it was shown in the last section. A numerical solution for $\delta$ can be obtained
from equation (\ref{delta extreme}). This gives $\delta_{l}$ as a function
of $l$, from which the function $\delta\left(x\right)$ can be obtained
by replacing $x=l/\omega$ and fitting the points. This procedure is done in figure \ref{fig9}, for $T=5$ K and $\mu=250$ meV, where $\delta$ was fitted
with a Boltzmann function. Notice that the figure \ref{fig9}(a) is in agreement
with the figure \ref{fig7}, where $\delta$ increases with $l$
and thus with $1/B=x=l/\omega$, and it tends to the limit
$\delta\sim5.3\times10^{-3}\unit{T^{-1}}$. Once obtained $\delta\left(x\right)$,
the positive envelope as a function of $x$ is given by equation (\ref{E+}),
while the negative envelope is $E_{-}=-E_{+}$. This can be seen in
figure \ref{fig9}(b), where the MO and the envelopes are shown. We consider
only the terms $m=1,2$ in the sum of equation (\ref{E+}), which
shows that only the few first terms are needed.

\section{Conclusions}

We analyzed the magnetic oscillations (MO) in pristine graphene, under a perpendicular magnetic field and taking into account the Zeeman effect. For a constant Fermi energy $\mu>0$ and zero temperature we showed that, due to the Zeeman effect, the MO consist of two sawtooth peaks, each corresponding to one spin. Both peaks have the same frequency, but different amplitude and phase. We obtaining that it requires high Fermi energy, about $\mu\gtrsim0.1$ eV.

At $T\neq0$ K, the temperature effect is usually considered by introducing a reduction factor. Nevertheless, the resulting infinite series cannot be evaluated, and can only be approximated for special cases like low $B$, which in turn difficult the observation of the MO. Hence, we took another route by going back to the grand potential. From this we obtained that the MO at low $T$ can be expressed as the MO at zero temperature, plus small correction functions. These functions are Fermi-Dirac like, each centered around the MO peaks at zero temperature. Moreover, they are very small unless the magnetic field is close to the corresponding peaks, which can be useful from a practical sense. Using this expression we then analyzed how the temperature affects the observation of the spin splitting in the MO. We show that, in order of magnitude, the observation is possible when the thermal energy $kT$ is less than the spin splitting energy $2\mu_{\mathrm{B}}B$. One would also expect that any kind of disorder would hide even further the spin splitting observation in the MO. 

We also analyzed the shift of the MO extrema as the temperature increases. We obtained an expression which was solved numerically, showing that the shift increases with the temperature and decreases with the magnetic field, implying that the MO are not anymore periodic at $T\neq0$. This behavior could be useful to measure temperature changes from the MO. For instance, one possibility would be a graphene device which measures the MO around a particular magnetic field. Then, by analyzing the extrema shift from its known value at zero temperature, one can infer the temperature.
Finally, we obtained an analytical expression for the MO envelope, which in turn depends on the shift of the extrema.  

\ack
This paper was partially supported by grants of CONICET (Argentina
National Research Council) and Universidad Nacional del Sur (UNS)
and by ANPCyT through PICT 2014-1351. Res. N 270/15. N: 2014-1351, and PIP 2014-2016. Res. N 5013/14. C\'odigo: 11220130100436CO research grant, as well as by SGCyT-UNS., J.
S. A. and P. J. are members of CONICET., F. E. acknowledge  research fellowship from this institution.

\appendix
\setcounter{section}{1}
\section*{Appendix}

We shall prove that equation (\ref{MT}) is in agreement with equation (\ref{Mt aprox}).
First of all, we recall the following properties for $x\in\mathbb{R}$:

\begin{equation}
\eqalign{\cot x =\tan\left(\pi/2-x\right) \cr
\arctan\left[\tan\left(x\right)\right]  =x-\pi\mathrm{floor}\left\{ \frac{x+\pi/2}{\pi}\right\}} ,\label{properties}
\end{equation}
where $\mathrm{floor}\left\{ \right\} $ is the floor function. Thus,
taking only one spin, we can write equation (\ref{MT}) as

\begin{equation}
\fl \frac{M_{T,s}}{A_{s}} =\frac{1}{2}-\omega\left(\frac{1}{B}+s\Delta\right)-\sum_{m}g_{m}\label{Mt aprox 4}
-\mathrm{floor}\left\{ 1-\omega\left(\frac{1}{B}+s\Delta\right)-\sum_{m}g_{m}\right\}, 
\end{equation}
where we have defined $g_{m}=\left[1+e^{-\beta\mu\left(B-B_{m}\right)/2B_{m}}\right]^{-1}$,
with $B_{m}^{-1}=n_m/\omega-s_m\Delta$. Considering $\varepsilon_{f}<\mu\leq\varepsilon_{f+1}$,
which implies $B_{f+1}\leq B<B_{f}$, we have 

\begin{equation}
\label{cases}
g_m=\cases{< 1/2&for $m \leq f$\\
	\geq 1/2&for $m\geq f+1$\\}
\end{equation}
Furthermore, at low temperatures $g_{m}\ll1$ if $m<f$, while $g_{m}\rightarrow1$ if $m>f+1$. These considerations, along with the properties
of the floor function, implies that for $B_{f+1}\leq B<B_{f}$,
equation (\ref{Mt aprox 4}) can be written as

\begin{eqnarray}
\fl\frac{M_{T,s}}{A_{s}} = \frac{1}{2}-\omega\left(\frac{1}{B}+s\Delta\right)-\mathrm{floor}\left\{ 1-\omega\left(\frac{1}{B}+s\Delta\right)\right\} \label{Mt aprox 5}\nonumber\\
-\sum_{m\leq f}\frac{1}{1+e^{-\beta\mu\left(B-B_{m}\right)/2B_{m}}}+\sum_{m\geq f+1}\frac{1}{1+e^{\beta\mu\left(B-B_{m}\right)/2B_{m}}}.
\end{eqnarray}
Consequently, from equation (\ref{M0}) and the properties given by equation
(\ref{properties}), equation (\ref{Mt aprox 5}) becomes

\begin{equation}
M_{T,s}=M_{s}-\sum_{m\leq f}\frac{A_{s}}{1+e^{-\beta\mu\left(B-B_{m}\right)/2B_{m}}}
+\sum_{m\geq f+1}\frac{A_{s}}{1+e^{\beta\mu\left(B-B_{m}\right)/2B_{m}}},
\end{equation}
so equation (\ref{MT}) effectively reduces to equation (\ref{Mt aprox})
if $\varepsilon_{f}<\mu\leq\varepsilon_{f+1}$.

\section*{References}
\bibliography{graphene}	

\end{document}